\documentclass[twocolumn, showpacs, preprintnumbers, amsmath, amssymb] {revtex4}
\usepackage{times}
\usepackage{amsmath}
\usepackage{amssymb}
\usepackage{graphicx}
\usepackage{natbib}
\usepackage{psfrag}
 \usepackage{setspace}
 \usepackage{color}

\renewcommand{\deg}{\mbox{$^\circ$}}

\begin{document}

\title{On the existence of stable seasonally varying Arctic sea ice in simple models}

\author{W. Moon$^{1}$, \& J. S. Wettlaufer$^{1,2}$  }

\affiliation{$^{1}$Yale University, New Haven, Connecticut 06520, USA\\$^{2}$NORDITA,  Roslagstullsbacken 23, SE-10691 Stockholm, Sweden}


\date{\today}

\begin{abstract}
Within the framework of lower order thermodynamic theories for the climatic evolution of Arctic sea ice we isolate the conditions required for the existence of stable seasonally-varying solutions, in which ice forms each winter and melts away each summer.  This is done by constructing a two-season model from the continuously evolving theory of \cite{EW09} and showing that seasonally-varying states are unstable under constant annual average short-wave radiative forcing.  However, dividing the summer season into two intervals (ice covered and ice free) provides sufficient freedom to stabilize seasonal ice.  Simple perturbation theory shows that the condition for stability is determined by when the ice vanishes in summer and hence the relative magnitudes of the summer heat flux over the ocean versus over the ice.  This scenario is examined within the context of greenhouse gas warming, as a function of which stability conditions are discerned.
\end{abstract}


\maketitle


\section{Introduction}
The recent rapid decline of the Arctic sea ice cover captures efforts to constrain cause, effect and sensitivity in the evolution of the state of the system  \citep[e.g.,][]{perovich2009, Kwok:2011}.  A principal focus is the evolution of the summer sea ice minimum; if the ice cover vanishes in summer, some argue that this constitutes an irreversible change akin to a saddle-node bifurcation (where two fixed points merge and annihilate), whereas others argue that such a so-called ``tipping point'' is not associated with this transition, which would then be reversible.  
The arguments involve a range of methods from theoretical treatments \citep{thorndike1992, EW09, MW:2011, Eisenman2012, Bjork:2012} and global climate model simulations \citep[e.g.,][]{holland2006, Winton:2008, Tietsche:2011}, to extrapolation of observations \citep[e.g.,][]{Serreze:2011, Stroeve:2012}.  

The heuristics of the ice-albedo feedback underlie the notion of an abrupt and irreversible transition from the perennial ice state to {\em either} a seasonal {\em or} ice-free sea ice state; progressive reduction of the ice cover continues due to the secular increase in the sensible heat of the low albedo ocean.  Here we seek to lay bare the specific conditions under which the ice-albedo feedback drives a transition to a seasonal state that is stable.  

\cite{thorndike1992} developed an analytical model coupling sea ice growth to climate by calculating the annual cycle in four stages across a warm and a cold season.  For each season he chose representative constant values of radiative fluxes and the albedos of sea ice and the ocean.  His climate forcing took the form of a constant poleward heat transport within a range of which he found two stable steady state solutions and one unstable steady state solution.  The perennial ice and ice-free states were stable and the seasonally-varying states, in which the ice vanishes each summer and returns each winter, were unstable.   The  origin of the instability was the albedo difference between the ice and the ocean.

 \cite{EW09} extended the approach of \cite{thorndike1992} to develop a single evolution equation for the state of the system forced by monthly observations of the radiative fluxes.   They modeled ice-albedo feedback by treating the sea ice albedo as a function of sea ice thickness, transitioning continuously from that of sea ice to that of the ocean.   In analogy with the poleward heat transport of \cite{thorndike1992}, they forced the system with an additional heat flux $\Delta F_0$, generically associated with greenhouse gas forcing, and analyzed the fixed points to their evolution equation.  Upon increase of $\Delta F_0$ they found that stable seasonally-varying states emerged continuously from the perennial ice states.  These stable seasonal ice states persisted for a range of $\Delta F_0$ and then were lost to a saddle-node bifurcation (where two fixed points merge and annihilate) leading to an ice free state.  While maintaining the time dependence of the forcing, \cite{Eisenman2012} further simplified the model of \cite{EW09} to show that over a wide range of parameter choices the model can, among three other scenarios, capture the behavior shown by \cite{thorndike1992}.   \cite{Eisenman2012} provides a thorough summary of the models and methods used to predict four general scenarios under which ice retreat may occur in a warming climate, all of which he was able to reproduce within the scope of his wide ranging parameter study. 

The detailed construction of such conceptual models underlies the distinction between the nature and stability of their fixed points.  The value of seeking the physical underpinning of this distinction is to focus thinking on the essential ingredients for such approaches to capture qualitatively distinct behaviors.  That is the goal of this paper.  In order to make the appropriate comparison with \cite{thorndike1992} we construct a two season model by averaging that of \cite{EW09}.  In order to make the paper reasonably self-contained we summarize the model of \cite{EW09} in the next section.  We then describe our partitioning of it into a two-season model after which we give the solutions in the different subseasons.  The seasonal solution is obtained by enforcing conservation of energy and mass to match the ice-covered and ice-free solutions across the subintervals of time.  We then find that the sufficient condition for stable seasonally-varying states resides in a finite time dependence of the short-wave radiative forcing.   While the quantitative nature of the stability depends on the quantitative nature of this time dependence, the generic structure does not.  

\section{Outline of Stability Argument}

We show that a minimal model that resolves the annual cycle into just two time intervals; a cold dark winter season and a warm sunny summer cannot produce a stable seasonal ice solution.  This is because such a two season model does not account for the fact that the  ice must vanish before the ocean can absorb heat,  by which time the solar radiative flux is smaller than in early summer.  However, breaking the summer season into two intervals provides sufficient freedom for a stable seasonal ice solution.  The first interval is ice covered and the second interval is ice free.  The argument proceeds by following the evolution of a positive perturbation to the system energy, which leads to thinner ice at the beginning of the summer.  Because energy is conserved, the thinner ice in the first summer interval insures this period is shorter and that the second interval is longer than either would be in absence of the perturbation.  In consequence, the amount of energy absorbed in the second summer interval is larger by an amount proportional to the ratio of the summer heat flux over ocean to the summer heat flux over ice.  If this ratio is positive, the perturbations grow as described by equation (\ref{eq:seasonalinst}), and the seasonal ice solution is unstable.   This scenario is shown schematically in Figure  \ref{fig:schematic} wherein one can see that the stability of seasonal ice depends on when in the summer the ice melts, a degree of freedom not available to a model with just two seasons. The reader familiar with \cite{thorndike1992}  and \cite{EW09} can follow the essential argument by reading Section \ref{sec:twoseason}.  Those interested in seeing behavior of the two-season model constructed from \cite{EW09} should begin with the next section.   

\section{Summary of Eisenman \& Wettlaufer (2009)\label{sec:twoseason}}

A two season model is constructed from \cite{EW09} (EW09) by dividing the year into a ``cold season'' and a ``warm season''.  To insure that the reader can follow this construction we summarize EW09 here. 

The state variable $E$ is the energy (with units W m$^{- 2}$ yr) stored in sea ice as latent heat when the ocean is ice-covered or in the ocean mixed layer as sensible heat when the ocean is ice-free, viz., 
\begin{equation}
E\equiv 
\begin{cases}
-L_i h_i & E<0 \textrm{  [sea ice]}\\
c_{ml}H_{ml}T_{ml} & E \ge 0 \textrm{   [ocean]} 
\end{cases},
\label{eq:sum0}
\end{equation}
where $L_i$ is the sea ice latent heat of fusion, $h_i$ its thickness, $c_{ml}$ is the specific heat capacity of the ocean mixed layer, $H_{ml}$ is its depth and $T_{ml}$ its temperature. Ignoring salinity effects,
the temperature $T(t,E)$ of the sea ice or ocean, determined by energy balance across the layer,  is measured relative to the freezing point ${T}_{m}$ as
\begin{align}
 T(t,E) = -{\cal{R}}\left[\frac{F_{D}(t)}{k_i L_i/E - F_T(t)}\right],  
 \label{eq:T}
\end{align}
where the ramp function is ${\cal R}(x \ge 0) = x$  and ${\cal R}(x<0) = 0$, the thermal conductivity of the ice is $k_i$, and the radiative flux quantity $F_{D}$ is discussed in detail immediately after equation \ref{eq:FD}, and $F_T(t)$ is described presently. 

In the same manner as in EW09 we linearize the Stefan-Boltzmann equation, $\sigma ({T})^4\approx \left(\sigma_0+\sigma_T \Delta T_i \right)$, where $\sigma$ is the Stefan-Boltzmann constant, $\Delta T$ is the deviation of the surface temperature $T(t,E)$ from the freezing point, $\sigma_0=316$Wm$^{-2}$ and $\sigma_T=3.9$Wm$^{-2}$K$^{-1}$ are chosen such that the equation is exact when ${T}=-30$\deg and when ${T}=0$\deg C (the approximate values of ${T}$ during most of the winter and summer, respectively). This allows us to express the temperature dependence of the outgoing long wave flux as $F_0(t) + F_T(t) T(t,E) $,
where $F_0(t)$ is $\sigma_0$ plus the specified sensible and latent heat fluxes from observation, and $F_T(t)=\sigma_T$.  An atmospheric model incorporating observations of Arctic cloudiness, atmospheric transport from lower latitudes and the meridional temperature gradient is used to determine the seasonally varying values of $F_0(t)$ and $F_T(t)$  \citep{EW09} but here we choose representative constant or seasonal values as described in Table 1.  

An essential aspect of the transitions we discuss is the nature of the ice albedo feedback.  Here, the Beer-Lambert law of exponential attenuation of radiative intensity with depth in a medium motivates a treatment of the dependence of the surface albedo with $E$ using a mixture formula with a characteristic ice thickness $h_{\alpha}$ for the extinction of shortwave radiation as 
\begin{align}
 \alpha(E)=\frac{\alpha_{ml}+\alpha_i}{2}+\frac{\alpha_{ml}-\alpha_i}{2} \text{tanh}\left(\frac{E}{L_i h_{\alpha}}\right).
 \label{eq:alpha}
\end{align}
This describes the fraction, $1-\alpha(E)$, of the incident shortwave radiation $F_S(t)$ absorbed by the ice.

The evolution of the state of the ice (or ocean) cover is determined by the balance of radiative and sensible heat fluxes at the upper surface, $F_{D} - F_T(t) T(t,E)$, the upward heat flux from the ocean $F_B$, and the fraction of ice exported from the domain $v_0 \mathcal{R}(-E)$ through a first order nonautonomous energy balance model as
\begin{align}
 \frac{dE}{dt}=f(t,E),
\label{eq:DS}
\end{align}
with
\begin{align}
 f(t,E)=F_{D}-F_T(t)T(t,E)+F_B+v_0 \mathcal{R}(-E), 
\end{align}
where 
\begin{equation}
F_{D}(t,E) \equiv  \left[1-\alpha(E)\right] F_S(t) -F_0(t) + \Delta F_0. 
\label{eq:FD}
\end{equation}
The term $F_{D} - F_T(t) T(t,E)$ is thought of as the difference between the incoming shortwave radiation at the surface $\left[1-\alpha(E)\right] F_S(t)$ and the outgoing longwave radiation
($\propto T^4$), augmented here by sensible and latent heat fluxes as described above and an additional amount associated with greenhouse gas forcing $\Delta F_0$.  Finally, we note that the ice export $v_0 \mathcal{R}(-E)$ is typically $\sim 10\%$ yr$^{-1}$, but the nonlinear relationship between ice thickness and ice growth rate highlights the possibility that in changing climates a time dependent value may be important in determining multiple ice states \citep[most recently see][and refs. therein]{Bjork:2012}.  In the two season model we neglect ice export.  

\section{A Two Season Model from EW09}

The measure for the division between the cold and warm seasons is taken to be the downwelling shortwave radiance $F_S (t)$, and we average the fluxes (all measured in Wm$^{-2}$) involved in the seasonal evolution of sea ice  over these seasons as 
\begin{align}
 F_S (t) &=
\begin{cases}
 0 & 0 \leq t \leq \frac{1}{2}~~ \text{~~~~~~Cold Season}\\
200  & \frac{1}{2} \leq t \leq 1~~ \text{~~~~~~Warm Season},
\end{cases}
\end{align}
 and
\begin{align}
 F_0(t) &=
\begin{cases}
 104  & 0 \leq t \leq \frac{1}{2} ~~ \text{~~~~~~Cold Season}\\
 64  & \frac{1}{2} \leq t \leq 1~~ \text{~~~~~~Warm Season},
 \end{cases}
\end{align}
wherein time is measured in years.  Because $F_T (t)$ 
does not change significantly over an entire season we take it  as  a constant 3.0\;Wm$^{-2}$K$^{-1}$.  
The form of the albedo that allows one to study the ice-albedo feedback is given by equation (\ref{eq:alpha}), which we return to in Section \ref{sec:I-A}, but in the two season treatment when we clearly have ice we use $\alpha_{i}$ and when 
we clearly have ocean we use $\alpha_{ml}$.   Other parameters and constants are provided in the Table. 

During the cold season the maximum $E_0$ decreases eventually reaching a minimum value $E_1$, which then increases to $E_0$ again during the warm season.  By imposing conservation of energy (and hence mass) and continuity across the seasonal transitions we determine the solutions $E(t)$ for the cold and warm seasons for the perennial ice, ice free and seasonal ice states in turn. 

\begin{widetext}
\begin{center}
\begin{table}
 \caption{Description and values of model parameters}
 \centering
\begin{tabular}{|l|c|c|}
  \hline
Symbol  & Description & Value \\
  \hline
$L_i$ & Latent heat of fusion of ice & $9.5\;\text{Wm}^{-3}\text{yr}$ \\
$C_{ml}H_{ml}$ & Ocean mixed layer heat capacity $\times$ depth & $6.3\;\text{Wm}^{-2}\text{yr K}^{-1}$ \\
$k_i$ & Thermal conductivity of ice & $2\;\text{Wm}^{-1}\text{K}^{-1}$ \\
$\alpha_i$ & Albedo of sea ice & $0.68$ \\
$\alpha_{ml}$ & Albedo of ocean & $0.2$ \\
$F_S$ & Incident shortwave radiation flux; Warm season & $200\;\text{Wm}^{-2}$ \\
$F_0$ & Temperature-independent surface flux; Cold season & $104\;\text{Wm}^{-2}$ \\
$\tilde{F}_0$ & Temperature-independent surface flux; Warm season & $64\;\text{Wm}^{-2}$ \\
$F_T$ & Temperature-dependent surface flux & $3.0\;\text{Wm}^{-1}\text{K}^{-1}$ \\
$F_B$ & Ocean heat flux & $2.0\;\text{Wm}^{-2}$ \\
$\Delta F_0$ & Imposed surface (Greenhouse) heat flux & $0 \text{~~to~~} 60 \;\text{Wm}^{-2}$ \\
$F_N$ & Temperature-independent excess heat flux; Cold season with sea ice& $ -102  \text{~~to~~} -42 \;\text{Wm}^{-2}$ \\
$\tilde{F}_N$ & Temperature-independent excess heat flux; Warm season with sea ice& $ 2 \text{~~to~~} 62 \;\text{Wm}^{-2}$ \\
$F^{*}_N$ &Temperature-independent excess heat flux; Warm season with ocean & $ 98  \text{~~to~~} 158 \;\text{Wm}^{-2}$ \\
$r$ & $F_T/C_{ml}H_{ml}$ & $0.48 \;\text{myr}^{-1}$ \\
$E_0$ & Maximum energy of sea ice during a year & Independent variable\\
$E_1$ & Minimum energy of sea ice during a year &  Independent variable\\
  \hline
\end{tabular}
\end{table} 
\end{center}
\end{widetext}

\subsection{Perennial Ice Solution}

\subsubsection{Cold Season}

The evolution equation during cold season is given by
\begin{align}
 \frac{dE}{dt}=~&-F_0 + \Delta F_0 + F_B - F_T \left(\frac{-F_0+\Delta F_0}{F_T-k_i L_i/E}\right)\\
 =~&\frac{k_i L_i F_{N}-F_B F_T E}{k_i L_i-F_T E},
\end{align}
where $F_{N}\equiv-F_0+\Delta F_0 + F_B$, which upon defining the constants
\begin{align}
 A=\frac{k_i L_i}{F_T}\qquad \text{and}\qquad B=\frac{k_i L_i F_{N}}{F_B F_T}, 
\end{align}
gives
\begin{align}
\left( \frac{A-E}{B-E}\right) dE = F_B dt.
\end{align}
We write the implicit form of the solution as 
\begin{align}
[E(t)-E_0]-(A-B)\text{ln}\left[1+\frac{E_0 - E(t)}{B-E_0}\right]=F_B dt, 
\end{align}
and  because $\mid B-E_0 \mid \gg \mid E_0-E(t) \mid$ in the perennial ice state, then 
\begin{align}
 [E(t)-E_0]\left[1+\frac{A-B}{B-E_0}\right] \approx F_B t,
\end{align}
which we rearrange as 
\begin{align}
 E(t)=~&E_0 + \left[\frac{F_B(B-E_0)}{A-E_0}\right]t \nonumber\\
 =~&E_0 + \left[\frac{k_i L_i F_{N}-F_B F_T E_0}{k_i L_i-F_T E_0}\right]t. 
\end{align}
This expresses the stabilizing energy balance during the cold season; the quantity multiplying time is negative with a magnitude that increases as
$E_0$ decreases thereby expressing the basic physics that thin ice grows faster than thick ice during winter. 
The cold season ends at $t$=$1/2$ when $E(t)$ reaches a minimum $E_1$ given by
\begin{align}
 E_1 = E_0 + \frac{k_i L_i F_{N}-F_B F_T E_0}{2(k_i L_i - F_T E_0)}.
 \label{eq:PImin}
\end{align}

\subsubsection{Warm Season}

During the warm season, the ice is ablating and hence $T(t,E)$ is zero, and the evolution equation is
\begin{align}
 \frac{dE}{dt}=(1-\alpha_i)F_S - \tilde{F}_0 +\Delta F_0 + F_B \equiv \tilde{F}_{N}, 
 \end{align}
which we integrate from $t=1/2$ to $1$ giving
\begin{align}
 E_0-E_1=\frac{1}{2}\tilde{F}_{N}.
\end{align}
With the use of equation (\ref{eq:PImin}) this can be expressed in terms of fluxes as
\begin{align}
 E_0=\frac{k_i L_i (F_{N}+\tilde{F}_{N})}{\tilde{F}_{N} F_T+F_B F_T}, 
\end{align}
from which we can see that the summer sea ice cover vanishes when $F_{N}+\tilde{F}_{N}$ = 0.
Therefore, this condition allows us to determine the greenhouse forcing $\Delta F_0$ associated with the vanishing of the ice cover viz., 
\begin{align}
 \Delta F_0 &= \frac{F_0 + \tilde{F}_0}{2}-(1-\alpha_i)\frac{F_S}{2}-F_B \nonumber \\ 
                    &\simeq53 ~ \text{W m}^{-2}.
\label{eq:vanish}
\end{align}
Such a simple expression demonstrates how the imbalance of heat fluxes over the Arctic Ocean is compensated for by the growth or decay of sea ice. 
The condition (\ref{eq:vanish}) for the vanishing of sea ice during summer represents a transition point in this energy balance. 

\subsection{Ice-Free Solution}
The other stable solution that we know exists as greenhouse forcing $\Delta F_0$ increases is ice-free \citep[e.g.,][]{thorndike1992, EW09}. We construct this solution in the two season setting as follows. 

\subsubsection{Cold Season}
The evolution equation is 
\begin{equation}
 \frac{dE}{dt}=-rE+F_{N},
 \label{eq:icefreecold}
\end{equation}
where $r=F_T/C_{ml}H_{ml}$ and again $F_{N}=-F_0+\Delta F_0 + F_B$.
The solution is thus
\begin{equation}
 E(t)=E_0 e^{-rt}+\frac{F_{N}}{r}\left(1-e^{-rt}\right), 
  \label{eq:icefreecoldsoln}
\end{equation}
and at the end of cold season, $E(t)$ reaches the minimum $E_1$, which is
\begin{equation}
 E_1 = E_0 e^{-\frac{1}{2}r}+\frac{F_{N}}{r}(1-e^{-\frac{1}{2}r} ).
\label{eq:IFmin}
\end{equation}

\subsubsection{Warm Season}

During the warm season, the shortwave radiative flux makes an important contribution to the system with 
the evolution equation taking the same form as equation (\ref{eq:icefreecold}) but with $F_{N}$ replaced by
$F^{\ast}_{N}\equiv(1-\alpha_{ml})F_S -\tilde{F}_0 + \Delta F_0 + F_B$ so that
\begin{align}
 \frac{dE}{dt}=-rE+F^{\ast}_{N}. 
\end{align}
Thus, the solution takes the form of equation (\ref{eq:icefreecoldsoln}) with $t \rightarrow t - 1/2$, viz., 
\begin{align}
 E(t)=E_1 e^{-r(t-1/2)}+\frac{F^{\ast}_{N}}{r}(1-e^{-r(t-1/2)}).  
 \end{align}
For a closed energy cycle $E(t)$ must reach $E_0$ at the end of the warm season giving
\begin{align}
 E_0 = E_1 e^{-\frac{1}{2}r}+\frac{F^{\ast}_{N}}{r}(1-e^{-\frac{1}{2}r}).
 \label{eq:IFmax}
\end{align}
Thus, equations (\ref{eq:IFmin}) and (\ref{eq:IFmax}) allow us to determine $E_0$ and $E_1$ for the ice free state which are
\begin{align}
 E_0 &= \frac{F_{N}e^{-\frac{1}{2}r}+F^{\ast}_{N}}{r(1+e^{-\frac{1}{2}r})}  \text{~~~and}\\
 E_1 &= \frac{F_{N}+F^{\ast}_{N}e^{-\frac{1}{2}r}}{r(1+e^{-\frac{1}{2}r})}.
\end{align}
Clearly the existence of ice free states requires that the minimum $E(t)$ must be greater than 0, which here implies 
$E_1 \geq 0$ and thus
\begin{align}
& F_{N}+F^{\ast}_{N} e^{-\frac{1}{2}r} \geq 0 \qquad \text{which is equivalent to} \nonumber\\
 & \Delta F_0 \geq \frac{F_0-F_B-e^{-\frac{1}{2}r}[(1-\alpha_{ml})F_S-\tilde{F}_0+F_B]}{1+e^{-\frac{1}{2}r}}\nonumber \\ 
                    &~~~~~~~\simeq~ 14 \text{W m}^{-2}.
\label{eq:minicefree}
\end{align}
We have thus captured analytically an essential feature of the bifurcation diagram of EW09 and the analysis of \cite{thorndike1992} of a hysteresis between perennial ice and ice free states. These two possible stable states exist over a range of $\Delta F_0$ determined by the minimum condition for ice free states, equation (\ref{eq:minicefree}), and the condition determining the value at which perennial ice vanishes, equation (\ref{eq:vanish}).  Now we examine the nature of the seasonal states. 

\subsection{Seasonal Ice Solution\label{sec:seasonal}}

\begin{figure}
\centering
\includegraphics[width=8.5cm]{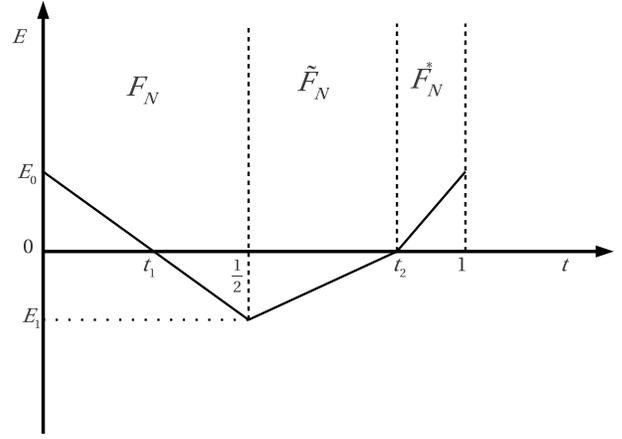}
\vspace{- 0.55 cm}
\caption{Seasonally-varying solutions in the two season model. Starting at $t=0$ from a positive  open ocean energy $E_0$ at the beginning of the cold season,  ice forms when $E(t)$ crosses into the 
negative region at $t=t_1$.  The ice grows during winter to reach the minimum $E_1$ at the end of the cold season, $t=1/2$. 
During the warm season, $E(t)$ becomes positive and the ice vanishes at $t=t_2$, after which  the ocean warms back to $E_0$.
The net radiative forcings $F_{N}$, $\tilde{F}_{N} $ and $F^{*}_{N} $ are described by equations (\ref{eq:cold}-\ref{eq:oceanwarm})}
\label{fig:seasonally}
\end{figure}

We seek to understand the conditions under which seasonal ice states are stable. 
The approach is simply to combine the perennial ice and ice free solutions. At the end of the warm season $E(t)$ reaches the positive maximum value $E_0$. 
Before the end of  the cold season sea ice must form such that the minimum $E_1$ is negative. 
Then, during the next warm season the ice ablates completely to return to $E_0$. 
Figure \ref{fig:seasonally} is a schematic of such seasonally 
varying solutions. Following the notation above we define 
\begin{align}
 F_{N} &=-F_0+\Delta F_0 + F_B,  &&\qquad\text{CS}  \label{eq:cold}\\ 
\tilde{F}_{N} &= (1-\alpha_i)F_S - \tilde{F}_0+\Delta F_0 + F_B, &&\qquad\text{IWS}   \label{eq:icewarm}\\
F^{*}_{N} &= (1-\alpha_{ml}) F_S-\tilde{F}_0 + \Delta F_0 + F_B, &&\qquad\text{IFWS}  \label{eq:oceanwarm}
\end{align}
where CS, IWS and IFWS denote Cold Season, Icy Warm Season and Ice Free Warm Season respectively. 
The nature of the seasonally varying solutions requires a crossover time $t_1$ from the ice free state to the ice covered state during the cold 
season, and a return to the ice free state at a time $t_2$ during the warm season. 
Thus, the solution for the seasonally-varying state must be solved incrementally in the  four stages shown in Figure  \ref{fig:seasonally}. 

\subsubsection{$0 \leq t \leq t_1$}
The cooling of the ocean is governed by the evolution equation for $E(t)$ for the ice free state which is
\begin{align}
 \frac{dE}{dt}=-rE+F_{N}.
\end{align}
We thus integrate with an integrating factor $e^{rt}$ from $t=0$ to $t=t_1$, noting that $E(t=t_1)$ = 0,  
to find
\begin{align}
 t_1=\frac{1}{r}\text{ln} \left(1-\frac{r}{F_{N}}E_0\right) \simeq -\frac{E_0}{F_{N}},
\end{align}
exploiting the fact that $\mid rE_0 |\mid \ll \mid F_{N} \mid$.

\subsubsection{$t_1 \leq t \leq 1/2$}
Once ice has formed it grows according to the evolution equation we used for the perennial ice state during the cold season;
\begin{align}
 \frac{dE}{dt}=\frac{k_i L_i F_{N}-F_B F_T E}{k_i L_i - F_T E}.
\end{align}
Integrating from $t=t_1$ to $t=1/2$ and using $A$ and $B$ from above we find 
\begin{align}
 (E_1-0)\left[1+\frac{A - B}{B - 0}\right]= F_B \left(\frac{1}{2}-t_1\right), 
\end{align}
and hence 
\begin{align}
 E_1 &=\frac{B}{A} F_B \left(\frac{1}{2}-t_1\right)
        =\left[\frac{k_i L_i F_{N}/F_B F_T}{k_i L_i/F_T} \right]F_B \left(\frac{1}{2}+\frac{E_0}{F_{N}}\right) \nonumber\\
        &= E_0 + \frac{1}{2} F_{N}. 
\label{eq:seasonalmin}
\end{align}

\subsubsection{$1/2 \leq t \leq t_2$}
Having passed through the minimum of $E(t)$ the system begins to warm.  It evolves with the albedo of sea ice and is described by
\begin{align}
 \frac{dE}{dt}=\tilde{F}_{N}, 
\end{align}
which we integrate from $t=1/2$ to $t=t_2$ to find
\begin{align}
 -E_1=\tilde{F}_{N} \left(t_2 - \frac{1}{2}\right),
\end{align}
from which, with the aid of equation (\ref{eq:seasonalmin}), we find $t_2$ as
\begin{align}
 t_2 = \frac{1}{2}-\frac{E_0}{\tilde{F}_{N}}-\frac{1}{2}\frac{F_{N}}{\tilde{F}_{N}}.
\end{align}

\subsubsection{$t_2 \leq t \leq 1$}

As $t$ passes through $t_2$ the sea ice vanishes and the radiatively exposed ocean warms according to
\begin{align}
 \frac{dE}{dt}&=-rE+F^{*}_{N} \nonumber \\
 &=-rE+(1-\alpha_{ml})F_S-\tilde{F}_0+\Delta F_0 + F_B, 
\end{align}
from which we find
\begin{align}
 E_0 &=\frac{F^{*}_{N}}{r}\left(1-\text{exp}\left[-~r\left(\frac{1}{2}+
             \frac{E_0}{\tilde{F}_{N}}+\frac{1}{2} \frac{F_{N}}{\tilde{F}_{N}}\right)\right]\right) \nonumber \\ 
  &\simeq \frac{1}{2}F^{*}_{N}+\frac{F^{*}_{N}}{\tilde{F}_{N}}E_0+\frac{1}{2}F^{*}_{N}
          \frac{F_{N}}{\tilde{F}_{N}} \nonumber \\
  &=\left[ \frac{\tilde{F}_{N}+F_{N}}{\tilde{F}_{N}-F^{*}_{N}} \right] \frac{F^{*}_{N}}{2}. 
\end{align}
wherein we rely on the observation that 
$r\left(\frac{1}{2}+\frac{E_0}{\tilde{F}_{N}}+\frac{1}{2} \frac{F_{N}}{\tilde{F}_{N}}\right)$ is small throughout the 
entire range of $\Delta F_0$ studied in EW09.  This originates in the fact that the term describing the excess longwave radiative flux due to the temperature change of the ocean, $-rE$, is small relative to the surface radiative flux, $F^{*}_{N}$.  However, as $\Delta F_0$ becomes very large the approximation breaks down.

\begin{figure}
\centering
\includegraphics[width=8.5cm]{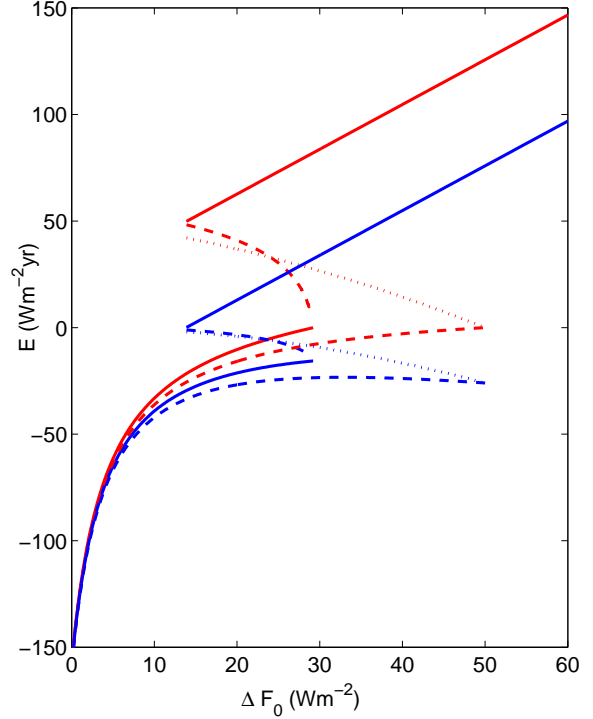}
\vspace{- 0.55 cm}
\caption{Bifurcation diagram for the two season model as a function of the external,  greenhouse gas forcing, heat flux $\Delta F_0$. 
Thick lines represent summer and winter energies calculated numerically and the 
dashed-dot lines the summer and winter energies calculated analytically. Here, seasonal ice vanishes in a saddle-node bifurcation.}
\label{fig:bifurcation}
\end{figure}

The expected behavior of the system is summarized in Figure 2. Dashed-dot lines give the behavior calculated analytically using the approach described above. Thick lines represent summer and winter energy values of 
the stable steady state solutions calculated numerically with the representative radiative flux values (the same as those in the analytical calculation) and ice export excluded.  
Clearly the bifurcation points occur at different values of $\Delta F_0$, which is due to the fact that we plot the analytical results using all of the approximations discussed, whereas these approximations are not made in the numerical calculations.  However, the point is that the main physical interpretation is the same.  We see in both cases the range of $\Delta F_0$ in which the two stable solutions (the perennial ice and ice free states) and one unstable solution (the seasonally varying state) coexist.  
Although this bifurcation diagram is sufficient to interpret the stability of steady state solutions parametrically, we seek an
expression defining the conditions for stability that will provide insight into the minimal physical constraints on the stable existence of seasonal ice. 

\section{Stability of seasonal ice and the time variation of shortwave radiative flux}

Having found the solutions for a two-season variant of EW09 we now examine their stability in the usual manner, taking each in turn.

\subsection{Perennial ice state}

First, we assess the stability of the perennial ice states by perturbing $E_1$ in equation (\ref{eq:PImin}) as ${E_1}^{\prime} \simeq E_1 + \delta^{\prime}$ which is accomplished by setting $E_0$ to $E_0+\delta$ such that $|\delta| \ll |E_0|$ giving 
\begin{align}
 {E_1}^{\prime} &=E_0+\delta+\frac{k_iL_i F_{N}-F_B F_T(E_0+\delta)}{2(k_i L_i-F_T(E_0+\delta))} \nonumber \\
                                 &\simeq E_0+\frac{k_i L_i F_{N}-F_B F_T E_0}{2(k_i L_i-F_T E_0)}
            +\left[1+\frac{F_T k_i L_i(F_{N}-F_B)}{2(k_i L_i-F_T E_0)^2}\right]\delta.
\end{align}
Recalling that 
\begin{align}
 E_0=\frac{k_iL_i(F_{N}+\tilde{F}_{N})}{F_T(\tilde{F}_{N}+F_B)},
\end{align}
we have
\begin{align}
 {E_1}^{\prime}=E_1+\delta\left[1-\frac{F_T(\tilde{F}_{N}+F_B)^2}{2k_iL_i(-F_{N}+F_B)}\right] = E_1 + \delta^{\prime}, 
\end{align}
and hence $\delta^\prime < \delta$ and the perturbation 
decays with time and that the perennial ice state solutions are stable.

\subsection{Ice free state}
Carrying out the same procedure with the ice-free solutions we find
\begin{align}
\delta^\prime = \delta e^{-r}.
\end{align}
Because $r$ is positive, the perturbation decays and the ice-free states are stable.

\subsection{Seasonally varying state}

When we perturb the initial energy $E_0$ of these states in the same manner we find
\begin{align}
 \delta^\prime= \delta \frac{F^{*}_{N}}{\tilde{F}_{N}},
\label{eq:seasonalinst}
\end{align}
where $F^{*}_{N}/\tilde{F}_{N} > 1$ and thus the perturbation grows with time. 
This is just a simple but more formal demonstration of what we understood from independent arguments; in this two season model the seasonally varying solutions are unstable. 
The physical origin of the inequality $F^{*}_{N} > \tilde{F}_{N}$ is the albedo difference between sea ice and water; for the same time periods and the same shortwave radiative flux, more energy is absorbed in the ocean than in the ice cover. 

\subsection{Varying the Shortwave Radiative Flux}

The main results will now be demonstrated.  The arguments leading to equation (\ref{eq:seasonalinst}) show that the instability of the seasonal ice states resides in the albedo difference between ice and water.  The warm season heat fluxes $\tilde{F}_{N}$ and $F^{*}_{N}$, given by equations (\ref{eq:icewarm}) and (\ref{eq:oceanwarm}), are the constant values during $1/2 \leq t \leq t_2$ and $t_2 \leq t \leq 1$ respectively. The sole difference between $\tilde{F}_{N}$ and $F^{*}_{N}$ resides in the albedos and hence the constant absorption of shortwave radiation in the ice and the ocean. Indeed, we emphasize that the shortwave, long wave and ocean heat fluxes are by prescription constants in the two season model.  We now consider relaxing this prescription by allowing for some temporal variation in the shortwave radiative flux $F_S(t)$ as
\begin{align}
 \langle \tilde{F}_{N}\rangle &\equiv\frac{1}{t_2-1/2} \int_{1/2}^{t_2} \tilde{F}_{N} \, dt \qquad\text{and} \nonumber \\
 \langle F^{*}_{N}\rangle &\equiv \frac{1}{1{~}-{~}t_2} \int_{t_2}^{1} F^{*}_{N} \,dt, 
 \label{eq:fluxave}
\end{align}
such that we can envisage that the average warm season values can possibly lead to stable conditions viz., 
\begin{align}
 \frac{\langle F^{*}_{N}\rangle}{\langle\tilde{F}_{N}\rangle} < 1.
 \label{eq:stabcond}
\end{align}

When perennial ice is present during the cold part of the seasonal ice cycle ($t_1 \leq t \leq 1/2$), 
the energy is governed by equation (\ref{eq:seasonalmin}) as
\begin{align}
 E_1=E_0+\frac{1}{2} F_{N}. \nonumber
\end{align}
Now, as the ice warms ($1/2 \leq t \leq t_2$) recall that the evolution equation is
\begin{align}
 \frac{dE}{dt}=(1-\alpha_i)F_S(t)-\tilde{F}_0 + \Delta F_0 +F_B, 
\end{align}
but here upon integration we maintain a time dependence in $F_S(t)$
\begin{align}
 0-E_1 = (1-\alpha_i)\int_{1/2}^{t_2} F_S(t) \, dt + (t_2 - \frac{1}{2})F,
\end{align}
where $F\equiv -\tilde{F}_0+\Delta F_0 +F_B$ and write ${\cal F}(t_f)\equiv\int_{t_i}^{t_f} F_S(t') \,dt'$,
which then leads to 
\begin{align}
 E_1= - (1-\alpha_i){\cal F}(t_2)-\left(t_2-\frac{1}{2}\right)F.
\end{align}
In the same manner, upon integration of the equation for the ice free system in $t_2 \leq t \leq 1$ we find
\begin{align}
 E_0 = (1-\alpha_{ml})[{\cal F}(1)-{\cal F}(t_2)]+(1-t_2)F.
\end{align}
Thus, energy conservation demands that we combine these two results according to equation (\ref{eq:seasonalmin})
to give
\begin{align}
 \frac{1}{2} F_{N}+(1-\alpha_{ml}){\cal F}(1)+F(1-t_2) \nonumber \\-(\alpha_i - \alpha_{ml}){\cal F}(t_2)+\left(t_2-\frac{1}{2}\right)F=0,
\end{align}
where we maintain the $t_2$ dependence for clarity in the development below.  First, however, we discuss the simplified version of the above expression 
\begin{align}
 {\cal F}(t_2)=\frac{\langle LW \rangle+(1-\alpha_{ml}){\cal F}(1)}{\alpha_i-\alpha_{ml}},
\label{eq:FluxBalance}
\end{align}
where $\langle LW \rangle\equiv\frac{1}{2}(F_{N}+F)$, which is the 
total outgoing longwave radiative flux during a year (including the ocean heat flux).  Whence, this is a statement of the fact that the net incoming shortwave radiative flux is balanced by outgoing longwave radiative flux, with a contribution from the ocean.
It is useful in many respects, one of which is that from it we can obtain the integrated incoming shortwave radiative flux into sea ice ${\cal F}(t_2)$.

Now, we reintroduce $t_2$ into equation (\ref{eq:FluxBalance}) in order to  construct an explicit version of equation (\ref{eq:stabcond}), the condition for the existence of stable seasonally varying states, which is 
\begin{align}
 \frac{(1-\alpha_{ml})}{(1-t_2)} [{\cal F}(1)-{\cal F}(t_2)] < \frac{(1-\alpha_i)}{t_2-1/2}~{\cal F}(t_2).
\end{align}
Upon use of the expression for ${\cal F}(t_2)$, we arrive at
\begin{align}
 t_2  <  \frac{1}{2} + \frac{(1-\alpha_i)[\langle LW \rangle 
         +(1-\alpha_{ml}){\cal F}(1)]}{2(\alpha_i-\alpha_{ml})(-\langle LW \rangle)}  \equiv  t^{*}.
\end{align}
Now we use the positivity of $F_S(t)$, and hence the fact that ${\cal F}(t)$ is a monotonically increasing function of its argument, to deduce the requirement  that 
\begin{align}
 {\cal F}(t^{*}) > {\cal F}(t_2),
\end{align}
which is equivalent to
\begin{align}
 {\cal F}(t^{*}) > \frac{\langle LW \rangle+(1-\alpha_{ml}){\cal F}(1)}{\alpha_i-\alpha_{ml}}.
 \label{eq:NandS}
\end{align} 
This is a sufficient condition for the existence of stable seasonally varying sea ice states.  In the discussion surrounding equation (\ref{eq:FluxBalance}) the 
right hand side was detailed as the integrated flux balance.  The inequality of equation (\ref{eq:NandS}) delivers the proviso that stable seasonal ice states exist under a restriction on the net shortwave radiance absorbed by sea ice during the warm season, which is clearly influenced by the albedo.  If the net radiative absorption  into the low albedo open ocean during the warm season is not too large, and equation (\ref{eq:stabcond}) is obeyed, then ice can still form the following cold season.  Over the entirety of the warm season (with and without ice) more net energy must be used to warm the ice cover and ablate it rather than be stored in the ocean, which imposes a time constraint. Indeed, it is evident that the position of $t_2$ during the warm season is determined by this inequality, and we examine this in the discussion surrounding Figures 3 and 4.  Several examples of shortwave radiances through the annual cycle are shown in Figure  3, from which we determine both ${\cal F}(t^{*})$ and ${\cal F}(t_2)$ 
and compare these as a function of $\Delta F_0$ in the Figure  4.  The green straight line shows ${\cal F}(t_2)$ which is to be compared with the curves of ${\cal F}(t^{*})$ for the different shortwave radiances from Figure  3.  We understand then from equation (\ref{eq:NandS}) that, for a given value of $\Delta F_0$, a {\em stable seasonally-varying state} must have ${\cal F}(t^{*})$ {\em above the green line} describing ${\cal F}(t_2)$.  
Clearly, the curves associated with the observational shortwave
radiance and the constant shortwave radiance (SW03) fall below the green line for this range of greenhouse forcing and thus describe unstable seasonal ice states.  However, the curves for SW01 and SW02 describe stable seasonal ice states from a wide range of  greenhouse forcing $\Delta F_0$.

\begin{figure}[!hbtp]
\centering
\includegraphics[width=8.5cm]{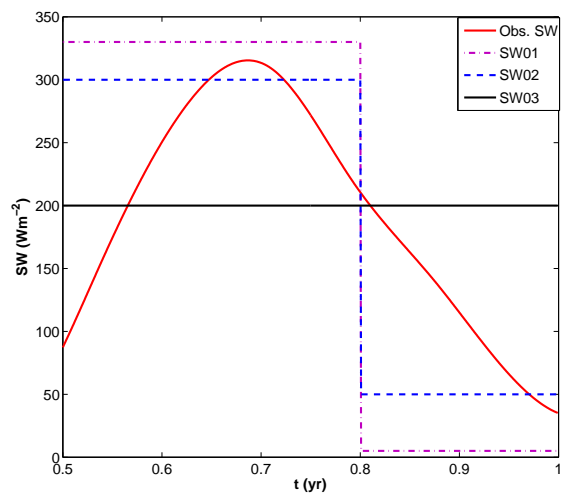}
\caption{Several short-wave radiance profiles. The red curve shows the short-wave radiance constructed from observational
data. The simplest treatments consistent with the model are two different constants divided at $t=0.8$, 
given by SW01 and SW02 and SW03 has no variation. All have an average of $200 \text{Wm}^{-2}$.}
\label{fig:fs}
\end{figure}
 
\begin{figure}[!hbtp]
\centering
\includegraphics[width=8.5cm]{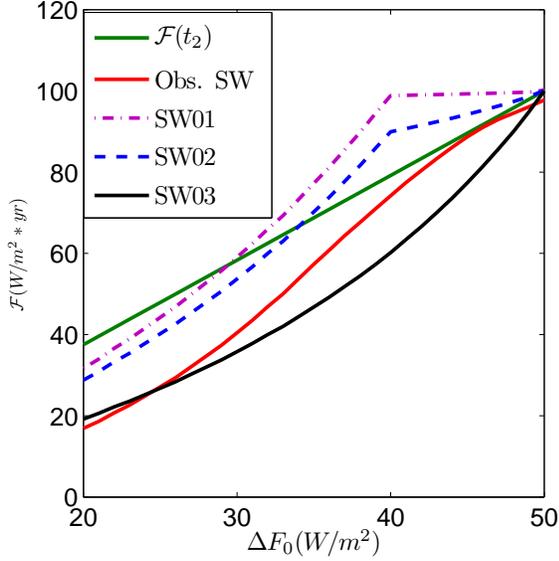}
\caption{${\cal F}(t_2)$ and ${\cal F}(t^{*})$ as a function of $\Delta F_0$. The green curve shows ${\cal F}(t_2)$ and the other curves are the different ${\cal F}(t^{*})$'s associated with the short-wave radiances of Figure  3. }
\label{fig:chi}
\end{figure}

\begin{figure}
 \includegraphics[height=5.5cm,width=7.5cm] {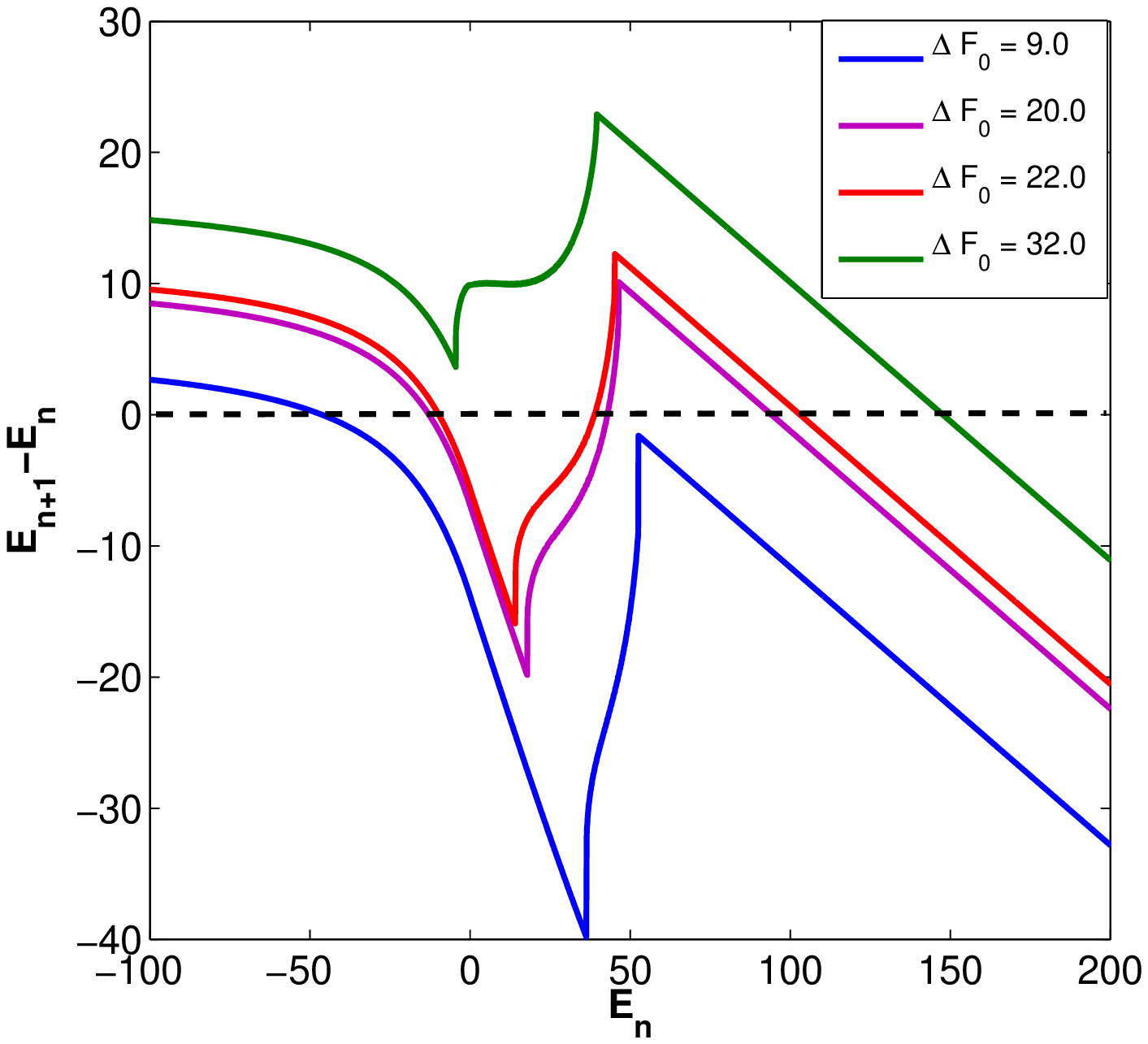}
\includegraphics[height=5.5cm,width=7.5cm] {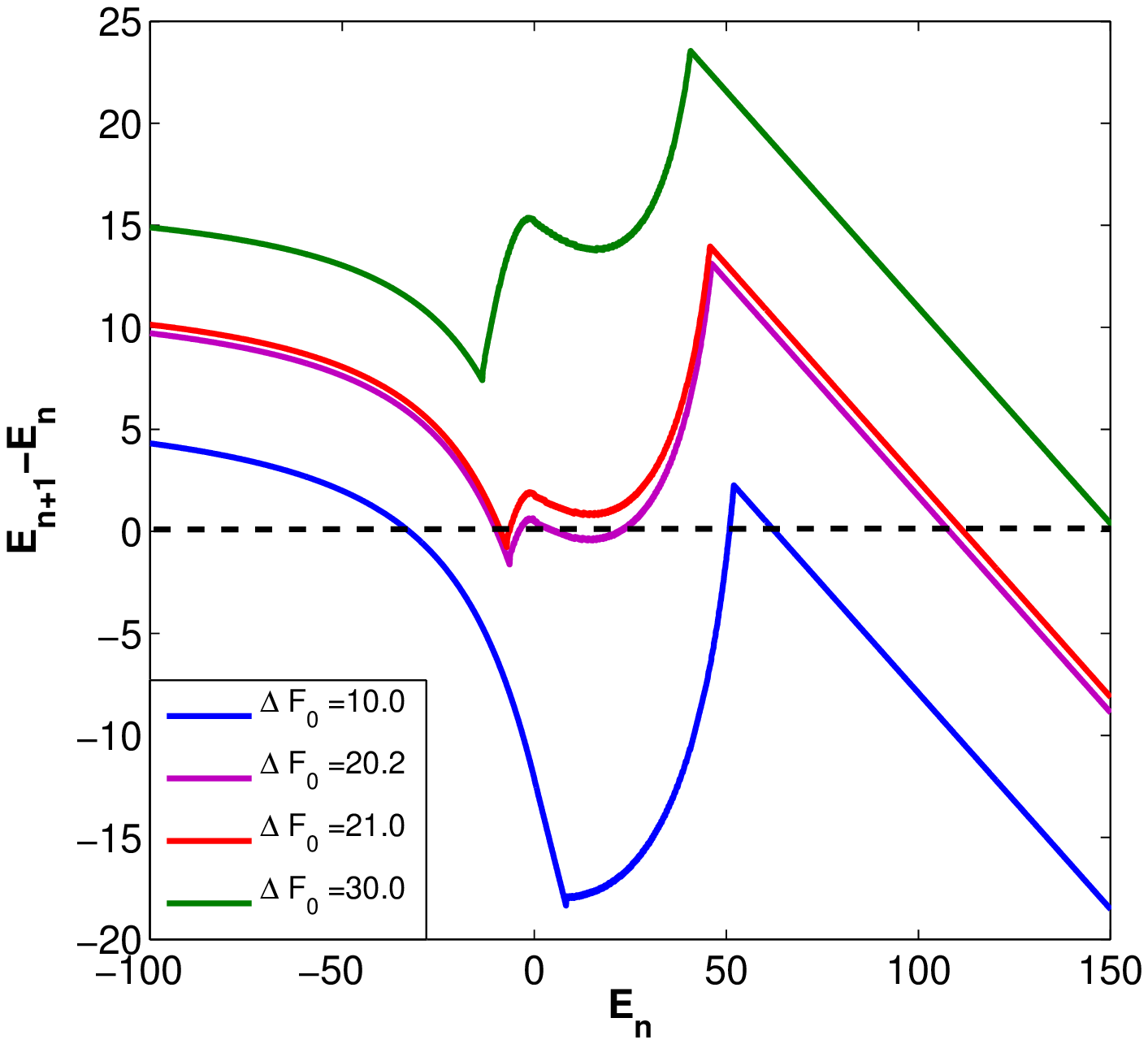}
\includegraphics[height=5.5cm,width=7.5cm] {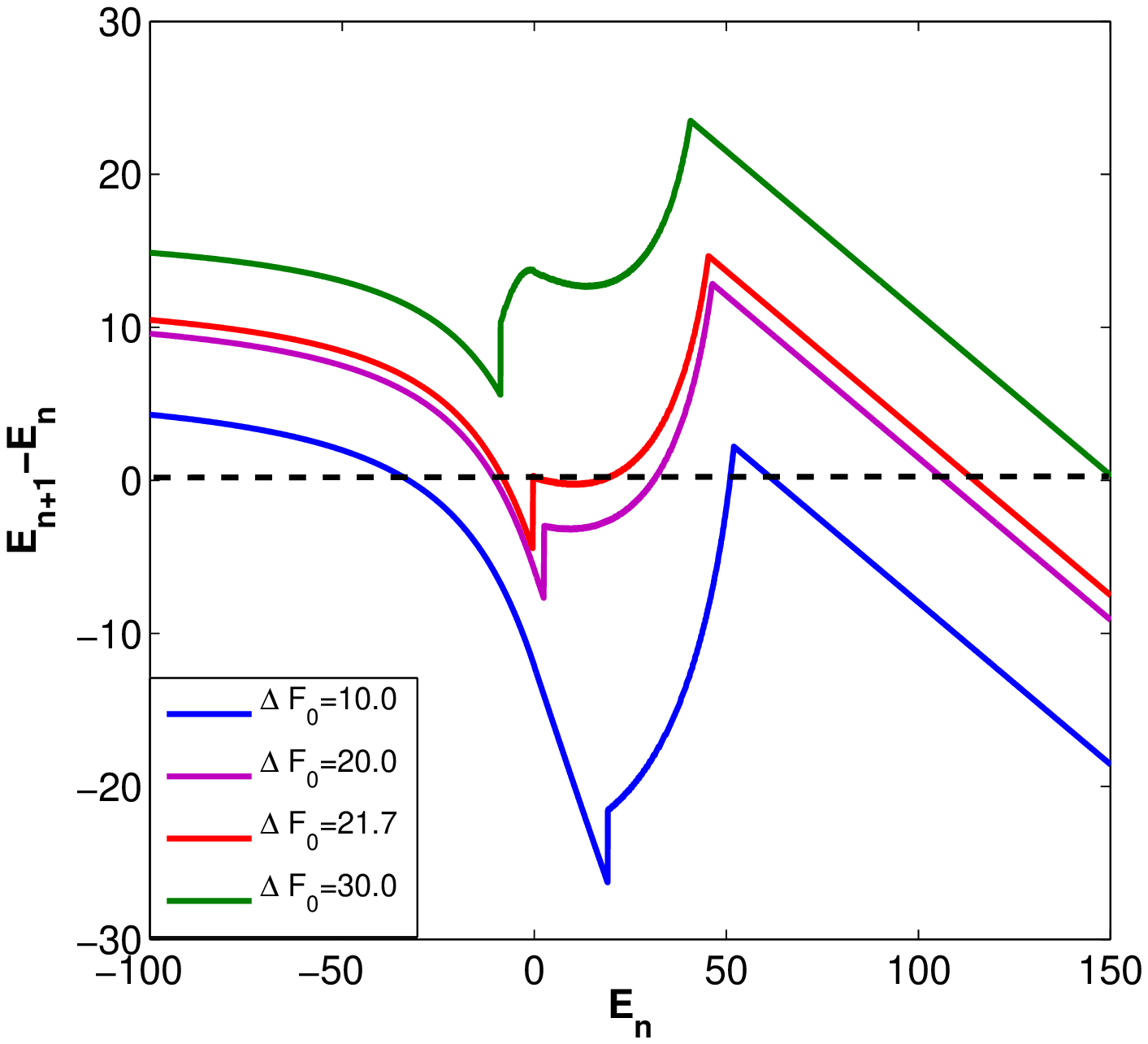}
\includegraphics[height=5.5cm,width=7.5cm] {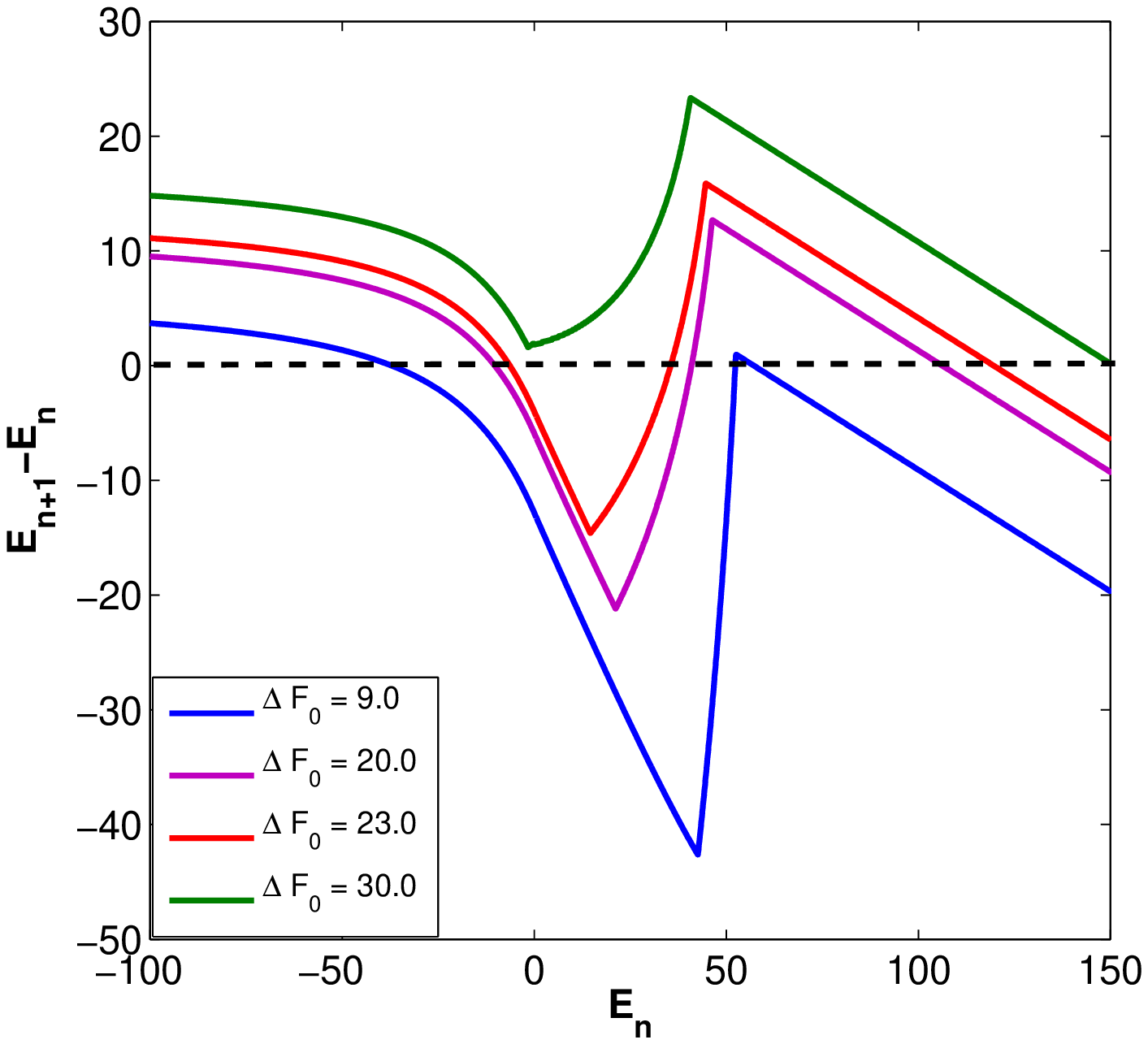}
\label{fig:poincare}
\caption{Poincar\'{e} sections of  energy (W m$^{- 2}$ yr) for various greenhouse forcing values $\Delta F_0$ and the short-wave radiances from Figure  3.  The observed (a), two examples of simple time dependence (b), (c) and a seasonally constant value (d).  
We take one of the summer energy values $E_n$ and then integrate for a year to get $E_{n+1}$. The steady state solutions are shown as the crossing points of the line $E_{n+1}$ - $E_{n}$ = 0. When the slope at these points is negative (positive), the solution is stable (unstable). }
\end{figure}

Fixing all fluxes except the shortwave radiance, we integrate the model numerically and 
examine the Poincar\'{e} sections to study  the nature and number of solutions in Figure  5.
Poincar\'{e} sections are developed and analyzed as follows.  
We take one of the summer energy values $E_n$ and then integrate for a year to get 
$E_{n+1}$. The points where $E_{n+1}-E_{n}$ = $0$ represent steady state solutions in the sense that they are periodic points of period unity. When the slope crossing zero  is negative (positive) the solutions are stable (unstable). In analogy with the results of \cite{thorndike1992}, for SW03, which takes a constant annual value,
we find only two stable steady states, the perennial ice state and ice-free state, and one unstable state, the seasonally-varying
state.    As expected from our stability analysis above, 
we find a stable seasonally-varying state with SW01 and SW02. 
Indeed, between the perennial ice and ice-free states, we find {\em  three} seasonally-varying states, 
one of which has negative slope and hence is stable. 

\begin{figure}
\includegraphics[width=8.5cm]{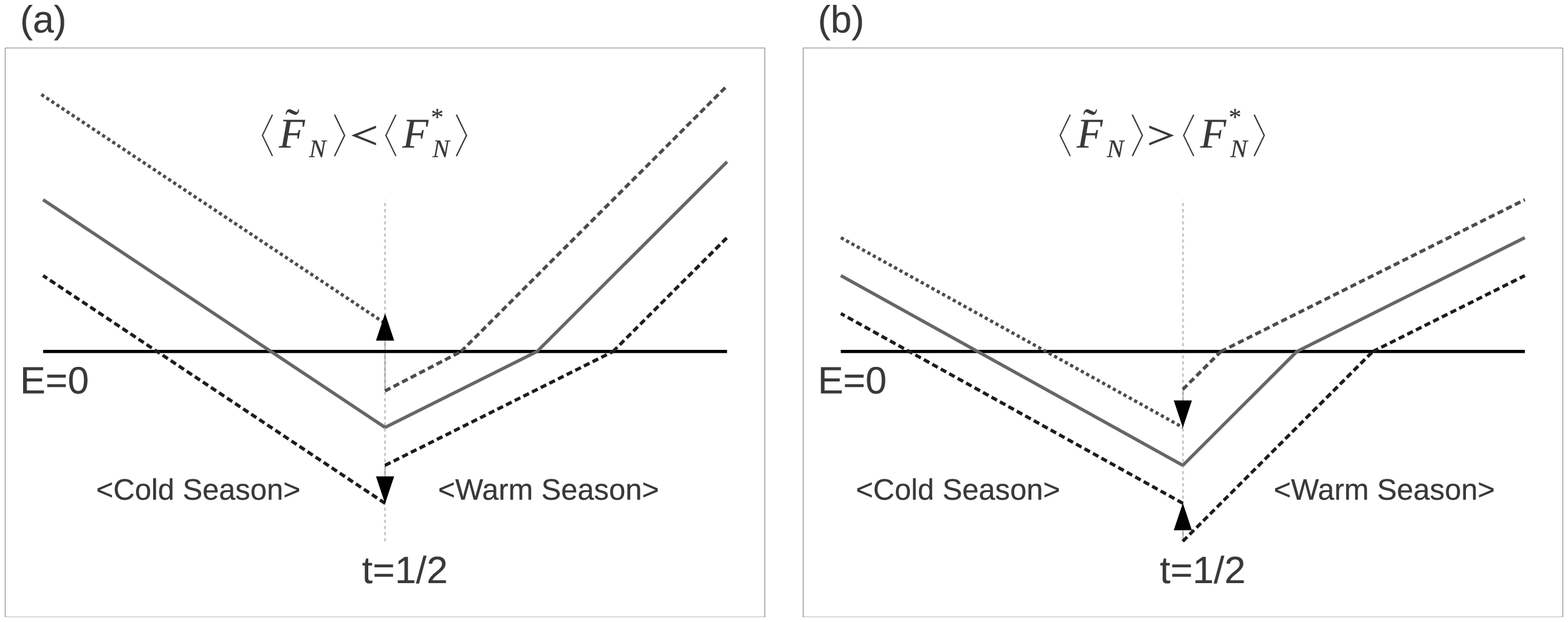}
\vspace{-1.75cm}
\caption{Schematic diagram of perturbation growth (decay) given to seasonally-varying solution. Both unstable (a) and stable (b) cases are shown. The thick lines denote steady state solutions and their slopes represent the average seasonal heat flux which depends on the presence or absence of sea ice. The dotted lines show an evolution of positively ($E > 0$) or negatively ($E < 0$) perturbed solutions starting 
from the beginning of the warm season $t=1/2$ during a given year. For the unstable case (a), the perturbed solution evolves with the same slope as the steady state solution, but after one cycle the deviation of the energy from the steady state solution is positive, as is represented by the direction of the arrows at $t=1/2$. Therefore, perturbations diverge from the steady state with time. 
Conversely, the perturbed solutions in the stable
case (b) converge to the steady state solution.}
\label{fig:schematic}
\end{figure}

Consider the unstable case for which $\langle \tilde{F}_{N} \rangle < \langle F^{*}_{N} \rangle$ is obeyed by
a perturbation to a steady seasonally-varying solution.  For a perturbation leading to a lower (more negative) winter energy value, there will be more sea ice at the end of winter.   Hence, there is more ice to be ablated the following summer and thus more survives as the system enters winter.  The feedback is positive.  Figure 6 is a schematic of the stability conditions and their manifestations. The slope represents the average total heat flux during the
 time period indicated thereby reflecting whether or not the stability condition  (\ref{eq:stabcond}) is satisfied. 
Therefore, the relative size of  $\langle F^{*}_{N} \rangle$ and $\langle \tilde{F}_{N} \rangle$ determines the stability of a solution.
  
\subsection{Sea Ice Albedo Feedback \label{sec:I-A}}

According to our analysis of the warm season equations, a time dependence of the shortwave radiance can underpin the 
stability of seasonally-varying sea ice states. Thus, it is worth emphasizing here that while an important role is played by the fixed albedo difference between ice and water in the two season model, there is
no sea ice albedo feedback.  Moreover, this is a zeroth order model that does not include an areal fraction of sea ice, and hence the fixed albedo difference cannot manifest itself as a feedback.  It was in this context that EW09 introduced the feedback formulation in equation (\ref{eq:alpha}). 

Now, we can use equation (\ref{eq:alpha}) and consider the ice albedo feedback under the assumption of a constant $F_S$
by suitable modification of equation (\ref{eq:fluxave}) as
\begin{align}
\langle \tilde{F}_{N} \rangle &=\frac{1}{t_2-1/2} \int_{1/2}^{t_2} \tilde{F}_{N} \, dt \nonumber \\ &=(1-\alpha_{AV})F_S-\tilde{F}_0+\Delta F_0+F_B,\nonumber \\
 \langle F^{*}_{N} \rangle &= \frac{1}{{~1}-{~t_2}} \int_{t_2}^{1} F^{*}_{N} \,dt  \nonumber \\ &= (1-\alpha_{ml})F_S-\tilde{F}_0+\Delta F_0+F_B,
\end{align}
 where $\alpha_{AV}$ is defined as
\begin{align}
 1-\alpha_{AV} = \frac{1}{t_2-1/2} \int_{1/2}^{t_2} \left[1-\alpha (E)\right] \, dt.
\end{align}
Thus,  $\alpha_{AV} < \alpha_i$, due to the ice thickness dependence (i.e., $E$ dependence) of the former, but it is still larger than $\alpha_{ml}$. Hence,
we see that even with the inclusion of the sea ice albedo feedback, 
it is impossible to generate stable seasonally-varying states so long as the shortwave radiative forcing $F_S$ is a constant as we will still have $\langle \tilde{F}_{N} \rangle < \langle F^{*}_{N} \rangle$. 

However, by including both ice albedo feedback and a time varying shortwave flux $F_S(t)$ we have
\begin{align}
&\langle \tilde{F}_{N} \rangle   \nonumber \\ & =\frac{1}{t_2-1/2} \int_{1/2}^{t_2} \left[1-\alpha (E)\right] \, F_S(t) dt -\tilde{F}_0+\Delta F_0+F_B,
\end{align}
and similar arguments as provided above upon replacement of  $\alpha_i$ by $\alpha_{AV}$ provide
\begin{align}
 \frac{(1-\alpha_i)\left[\left<LW\right>+(1-\alpha_{ml}){\cal F}(1)\right]}{2(\alpha_i-\alpha_{ml})(-\left<LW\right>)}&< \nonumber \\
 \frac{(1-\alpha_{AV})\left[\left<LW\right>+(1-\alpha_{ml}){\cal F}(1)\right]}{2(\alpha_{AV}-\alpha_{ml})(-\left<LW\right>)}.
 \label{eq:albedoav}
\end{align}
Therefore, for a given $\Delta F_0$, the inclusion of sea ice albedo feedback reduces the magnitude of the stability requirement for seasonally-varying states.

\section{Conclusion}

From the perspective of simple theoretical models we have asked whether the vanishing of summer Arctic sea ice cover would indicate an irreversible bifurcation in the climate system.  To this end we have analyzed in some detail the structure of a low order two season model describing the interaction of sea ice with the climate.  
The model was constructed by suitable simplification of the more complex single column treatment of \cite{EW09}, which is a non-autonomous system continuously forced by radiation climatology and other observed or inferred fluxes.  The reason for this approach is that they found a smooth transition from stable perennial ice states to stable seasonally-varying states as greenhouse forcing ($\Delta F_0$) increased.   Hence, they concluded that the loss of summer sea ice is not  irreversible; there is no hysteresis.  (Note however, the transition to an ice-free state is indeed a bifurcation of the saddle node type; exhibiting substantial hysteresis with the perennial ice state). On the other hand, \cite{thorndike1992} considered a two season theory forced by constant values and found no stable seasonal ice.   
Therefore, we sought to understand the minimal physical conditions for the existence of stable seasonal ice and focused our search on the basic difference between two season and continuously forced theories that could lead to robust qualitative distinctions in behavior. Such analysis is not possible with more complex models. 

First,  we showed that a model in which the annual cycle has only two time intervals cannot produce a stable seasonal ice solution.  Physically this is because such an approach does not account for the fact that the  ice must vanish before the ocean can absorb heat,  by which time the solar radiative flux is smaller than in early summer.  Sufficient freedom to stabilize seasonal ice is found by breaking the summer season into two intervals; ice covered and ice free.  The argument proceeds by following the evolution of (say) a positive perturbation to the system energy, which leads to thinner ice at the beginning of the ice covered summer.  Energy conservation insures that the thinner ice in the first summer interval shortens this period and that the second interval is longer than either would be in absence of the perturbation.  Therefore, the amount of energy absorbed in the second summer interval is larger by an amount proportional to the ratio of the summer heat flux over ocean to the summer heat flux over ice.  If this ratio is positive, the perturbations grow as described by equation (\ref{eq:seasonalinst}), and the seasonal ice solution is unstable.   The imbalance driving the instability resides in the ice-ocean albedo difference responsible for more energy being absorbed in the ocean than in the ice cover, insuring that the subsequent sensible heat of the ocean is too large to allow ice to form the following winter.  While we have not examined all physically realistic two season models it does not appear  possible to stabilize season ice with constant forcing, as shown in Figures 3 and 4.
The overall scenario is shown schematically in Figure  \ref{fig:schematic} wherein one can see that the stability of seasonal ice depends on {\em when} in the summer the ice melts; a degree of freedom not available to a model with just two seasons.

Second, the approach allowed us to capture analytically an essential feature of the bifurcation diagram of EW09 and the analysis of \cite{thorndike1992}, both of which showed a hysteresis between perennial ice and ice free states. We found that these two possible stable states exist over a range of greenhouse forcing. Simple expressions were found for the value of $\Delta F_0$ beyond which summer ice would vanish (equation \ref{eq:vanish}), and below which ice free conditions can no longer persist (equation \ref{eq:minicefree}).  

Third, the form of equation (\ref{eq:seasonalinst}) suggests that the simplest means of stabilizing seasonal ice states is to introduce a seasonality of the shortwave radiance $F_S(t)$.  This led to the ansatz (equation \ref{eq:stabcond}) suggesting that during the warm season reversing the ratio of the radiative balance between the ice covered and ice free states under {\em transient forcing} can have a stabilizing influence on seasonal ice states.  Pursuing this ansatz lead to an expression (equation \ref{eq:NandS}) describing the sufficient flux conditions for such stable states.   Simple examples are given in Figures 3 and 4, demonstrated numerically in Figure  5 and shown schematically in Figure  6. 

Fourth, two interesting results were found upon inclusion of the sea ice albedo feedback as characterized by equation (\ref{eq:alpha}).  We determined that the stability condition (\ref{eq:NandS}) is made less stringent by the implicit seasonal dependence of the albedo of the ice-ocean system, as shown in equation (\ref{eq:albedoav}).  Moreover, despite the presence of the sea ice albedo feedback, 
it is impossible to generate stable seasonally-varying states if the shortwave radiative forcing $F_S$ is a constant.  This codifies the generally understood importance in the real system of the timing of ice advance and retreat relative to solar insulation. 

Finally,  it is of broad interest to understand the nature of ice decay in the Arctic.  It is extremely difficult to use comprehensive GCMs to understand the qualitative distinction between the range of scenarios proposed \citep[see][and refs. therein]{Eisenman2012}, and extrapolation of observations on seasonal time scales is unwise \cite[][]{AMW:2012}.   The simplest theoretical approaches in this field began with the two season model of \cite{thorndike1992}, which predicts that once summer ice vanishes there is an irreversible change to the ice free state.  Whereas, EW09 and variants of continuously forced simple models suggest that such a change is reversible.  Having found the origin of this distinction, residing in the time variation of shortwave radiative forcing during the warm season, our analysis provides a simple and accessible framework to examine leading order effects on the nature and number of qualitative transitions in the state of the ice cover.  As such it is complimentary to the reduced version of EW09 studied by \cite{Eisenman2012} and, in the spirit of \cite{thorndike1992}, it provides a number of simple expressions trivial to use in the examination of the sensitivities of qualitative transitions.

\begin{acknowledgments}
WM thanks NASA for a graduate fellowship. JSW thanks the Wenner-Gren and John Simon Guggenheim Foundations, the Swedish Research Council and Yale University for support.  The comments, criticism and encouragement of N. Untersteiner at the beginning of this project shaped the final version of this paper. 
\end{acknowledgments}

\end{document}